\documentclass{osa-article}

\usepackage{amsmath}
\usepackage{algorithm}
\usepackage[noend]{algpseudocode}
\journal{boe}

\articletype{Research Article}

\begin{document}

\title{Computational multifocal microscopy}

\author{Kuan He,\authormark{1,*} Zihao Wang,\authormark{1} Xiang Huang,\authormark{2} Xiaolei Wang,\authormark{3} Seunghwan Yoo,\authormark{1} Pablo Ruiz,\authormark{1} Itay Gdor,\authormark{3} Alan Selewa,\authormark{3} Nicola J. Ferrier,\authormark{2} Norbert Scherer,\authormark{3,4} Mark Hereld,\authormark{2} Aggelos K. Katsaggelos,\authormark{1} and Oliver Cossairt\authormark{1}}

\address{\authormark{1}Department of Electrical Engineering and Computer Science, Northwestern University, Evanston, IL 60208, USA\\
\authormark{2}Mathematics and Computer Science, Argonne National Laboratory, 9700 South Cass Avenue, Lemont, IL 60439, USA\\
\authormark{3}Department of Chemistry, The University of Chicago, 5801 South Ellis Avenue, Chicago, IL 60637, USA\\
\authormark{4}James Franck Institute, The University of Chicago, 5801 South Ellis Avenue, Chicago, IL 60637, USA}

\email{\authormark{*}hekuan@u.northwestern.edu} 



\begin{abstract}
Despite recent advances, high performance single-shot 3D microscopy remains an elusive task. By introducing designed diffractive optical elements (DOEs), one is capable of converting a microscope into a 3D "kaleidoscope", in which case the snapshot image consists of an array of tiles and each tile focuses on different depths. However, the acquired multifocal microscopic (MFM) image suffers from multiple sources of degradation, which prevents MFM from further applications. We propose a unifying computational framework which simplifies the imaging system and achieves 3D reconstruction via computation. Our optical configuration omits chromatic correction grating and redesigns the multifocal grating to enlarge the tracking area. Our proposed setup features only one single grating in addition to a regular microscope. The aberration correction, along with Poisson and background denoising, are incorporated in our deconvolution-based fully-automated algorithm, which requires no empirical parameter-tuning. In experiments, we achieve the spatial resolutions of $0.35$um (lateral) and $0.5$um (axial), which are comparable to the resolution that can be achieved with confocal deconvolution microscopy. We demonstrate a 3D video of moving bacteria recorded at $25$ frames per second using our proposed computational multifocal microscopy technique. 
\end{abstract}

\section{Introduction}
\label{introduction}
Typically, an optical microscope focuses on one focal plane at a time. Thus, recovering 3D information with sufficient resolution usually requires sequentially z-scanning the focal planes. This process is time consuming and requires precise mechanical control. The long acquisition time of focal scanning microscopy fundamentally prevents the applications in \emph{in vivo} imaging, especially when the 3D movement of biomedical objects is desired. To realize \emph{single-shot} 3D microscopy, a standard microscope has to be modified in such a way that 3D information can be encoded onto a 2D plane. 

\begin{figure}
  \includegraphics[width=\linewidth]{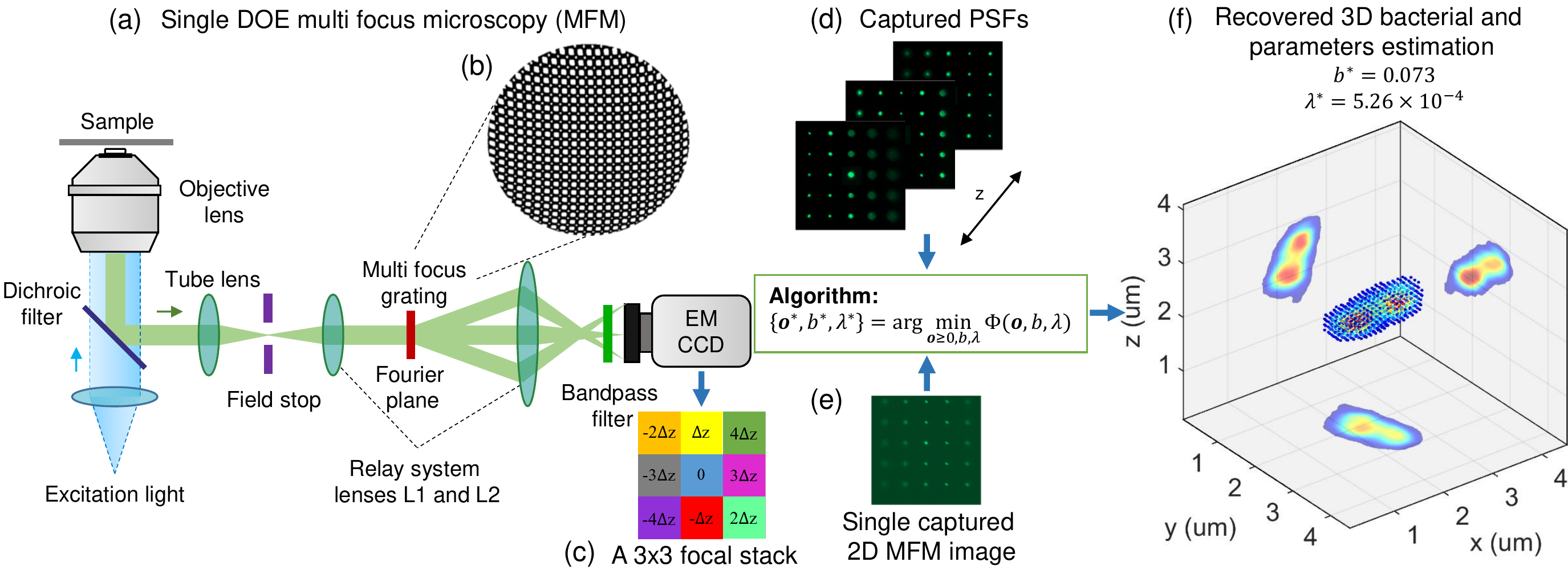}
  \caption{Single-DOE multifocal microscopy (MFM) setup (a) and computational 3D reconstruction pipeline (d-f). In (a), a conventional microscope is augmented by a $4f$ system. An MFG (b) is inserted at the Fourier plane of the $4f$ system to produce an array of $l \times l$ differently focused tile images in a single exposure (c; $l=3$). Note that our CMFM system discards CA corrective optics, significantly reducing the system complexity and cost compared to conventional MFM. We correct for CA computationally rather than optically. (d-f) The pipeline of proposed computational framework: by capturing a z-stack 3D PSFs (d), the algorithm can simultaneously recover the background noise $b^*$, the optimal regularizer parameter $\lambda^*$ and a high resolution 3D image (f) from a single captured 2D MFM image (e).}
  \label{fig:overview}
\end{figure}

One of the few modifications to achieve single-shot 3D microscopy, for example, is by placing a diffractive optical element (DOE) in the Fourier plane of a standard microscope (Fig. \ref{fig:overview}a). The DOE is designed as a distorted phase gratings \cite{Blanchard:99,Abrahamsson:16,Oudjedi:16}, also referred to as a multifocal gratings (MFG). The MFG projects different depth layers in a 3D volume onto different sub-regions of the imaging plane simultaneously. In particular, if the imaging plane is divided into $l \times l$ tiles and each tile is focused on one focal plane by design, then a 3D volume of $l^2$ depths can be retrieved by z-stacking all the tiles. An $l = 3$ case is shown in Fig. \ref{fig:overview}(c). This imaging technique is also referred as multifocal microscopy (MFM) \cite{Abrahamsson:13}.

MFM stands out among other single-shot 3D microscopy, e.g. light fields microscopy \cite{Prevedel:14, Pegard:16} and lensless 3D coded microscopy \cite{Adams:17, Coskun:10, Antipa:18}, for its capability of achieving diffraction-limited 3D resolution comparable to conventional focal scanning microscopy. However, as an imaging system, MFM faces several imaging issues, some of which are even more severe that prevents MFM from broader applications:
\begin{itemize}
\item \textbf{Out-of-focus blur.} 
MFM captures an array of multifocal planes. Like standard wide-field microscopy, each focal plane image is convolved by out-of-focus light. Estimating and removing out-of-focus blur requires implementation of a 3D deconvolution algorithm.
\emph{Our proposed algorithm builds on the regularized Rechardson-Lucy (RL) deconvolution, and features automatic estimation of the background noise and the optimal regularizer parameter.}
\item \textbf{Chromatic aberration (CA).} Conventionally, CA is optically corrected by adding a chromatic correction gratings (CCG) and a multifacet prism after the MFG \cite{Abrahamsson:13}, resulting in excessive hardware cost. In our work, we propose a simplified version by abandoning the CCG and the prism. Instead, we place a $10$ nm bandpass filter in front of the detector to mitigate CA, as shown in Fig. \ref{fig:overview}(a). We demonstrate that the resolution loss due to CA can be \emph{computationally} compensated. This comes from a fact that CA from MFG is directional due to tile geometry. In each tile, the PSF is stretched mainly in one direction, but preserves imaging quality in the orthogonal direction. \emph{Therefore, by jointly incorporating all the PSFs in each tile, our proposed 3D deconvolution algorithm enables restoring the resolution loss due to CA.} 
\item \textbf{Poisson noise due to limited photon budget.} There are three factors contribute to the limited photon budget of the MFM. First, as mentioned before, a narrower bandpass filter of $10$nm is used to mitigate CA, and thus, the total number of emission photons arriving at the detector is smaller than when using CA corrective optics. Second, the theoretical maximum efficiency of the MFG is smaller than one. For example, the theoretical limits are $68\%$ for a binary phase-only ($0$ or $\pi$) MFG with $3\times3$ diffraction orders, and $78\%$ for a MFG with $5\times5$ diffraction orders, which means $32\%$ or $22\%$ of the total emission light will be lost and cannot be collected by the detector. The efficiency can be improved by using a multi-phase MFG \cite{Mait:95} at the expense of the complicate fabrication process. Third, each tile only shares about $1/l^2$ of the total photons. The limited photon budget requires a noise modeling of Poisson process. \emph{We incorporate the Poisson noise modeling for MFM, for the first time.}

\item \textbf{Field-of-view (FOV).} As described above, MFM trades lateral FOV for depth information. Conventional MFM designs use large tile spacings for large objects. We address here that this design is not optimal for tracking dynamic objects, which are usually small in size but require large FOV. We analyze the FOV performance theoretically, and demonstrate that \emph{by modifying the MFG design to create a smaller tile spacing, small objects (e.g. bacteria) can be tracked over a larger lateral area than the tile width.}


\end{itemize} 

\subsection{Related works}
The 2D sensing limitation imposed by optical sensors makes 3D imaging an interesting and active research topic both in macroscopic and microscopic regime. In microscopy, several approaches have been proposed for a variety of imaging conditions. One scheme for achieving 3D imaging is to use coherent illumination, e.g. digital holographic microscopy \cite{pedrini2001short,schilling1997three,wang20174d}, variably-controlled light-emitting diodes (LED) arrays \cite{tian20143d,tian20153d}, etc. Yet the coherent modeling of light propagation does not apply to incoherent/fluorescent objects. For the incoherent case, one seminal idea is to introduce self-interference by projecting a set of Fresnel patterns \cite{rosen2007digital,rosen2008non,cossairt2016compressive}. However, most of the existing methods, especially for incoherent cases, require multiple exposures and massive processing time. This limitation prevents broader applications such as \emph{in vivo} imaging as the motion of the objects during capture process is hard to circumvent and thus, deteriorates the image quality \cite{ryu2017subsampled,wang2017dictionary}. Recent works have proposed to reduce the measuring requirement by computationally exploiting spatial-temporal redundancy of the scenes \cite{liu2017multiplexed, wang2017compressive}. Here, we review two types of computational microscopy that enable single-shot 3D fluorescence imaging.

\emph{Light field microscopy (LFM).} By introducing a microlens array on the primary image plane of a microscopy, LFM encodes 4D (spatial-angular) information into a 2D image and computationally recover the 3D objects \cite{Prevedel:14, Pegard:16}. Because LFM trades spatial resolution for single-shot capture, its spatial resolution is lower than that of the conventional microscope. Recently, a RL deconvolution method based on the wave optics theory \cite{Broxton:13} is derived and demonstrated to improve the resolution for LFM. So far, the best resolutions experimentally achieved by LFM are $\sim1.4$um and $2.6$um \cite{Prevedel:14} in the lateral and axial dimensions respectively. In addition, due to the variation on sampling density, the lateral resolution decreases to 
$\sim3.75$um at the focal plane \cite{Prevedel:14}. 

\emph{Lensless 3D microscopy.} Traditionally, lens-based cameras/microscopes map a point in the scene to a pixel on the detector. Lensless imaging architectures, instead, replace the lens with other encoding elements such that a point is mapped to many points on the detector and thus, require computation to recover. Several seminal literatures have proposed to place a single encoding element, such as a coded mask \cite{Adams:17} or diffuser \cite{Antipa:18}, directly in front of a detector. Therefore, the imaging system is compact and cost-effective, and has large FOV. A computational algorithm is then designed to make use of the physical effects, e.g. the Point Spread Functions (PSF), to recover the 3D information of the scene. However, the spatial resolution of lensless 3D microscopes is restricted to the pixel pitch of the detector. In particular, a lens-free fluorescent imaging platform was reported to achieve the spatial resolution of $\sim10$um based on a compressive sensing algorithm for sparse objects \cite{Coskun:10}. The recovered 3D volume consisted of two or three depth layers with intervals of $50$um or $100$um. 


\subsection{Computational multifocal microscopy} 
Our computational imaging framework, which we call computational multifocal microscopy (CMFM), balances and optimizes the processing capabilities of optics and computation. We demonstrate a CMFM system [Fig.\ref{fig:overview}(a)] that utilizes simplified optics, as well as algorithms that correct for CA and low-photon counts. The pipeline of our computational framework is shown in Fig.\ref{fig:overview}(d-f). 

We perform two experiments to demonstrate the effectiveness of our CMFM system. First, we capture a static image of several frozen periplasms in 3D to demonstrate the spatial resolutions of $0.35$um and $0.5$um in the lateral and axial dimensions [see Fig. \ref{periplasm reconstruction}(d)], which are comparable to those achievable with confocal deconvolution microscopy [see Fig. \ref{periplasm reconstruction}(b)]. Note that we compare $0.5$s captures with our CMFM instrument to a $20$s confocal scan taken with a dual spinning disk confocal microscope (Model: CSU-W1) made by Yokogawa Electric Corporation. Our CMFM results show similar 3D image quality, but achieve a 40x reduction in acquisition time. \emph{However, we kindly remind readers that the principle of the axial resolution improvement is different for the two techniques; Further discussion will appear in section \ref{Static scene}}. Second, we record a video of \emph{in vivo} bacterium at $25$ frames per second (see \textcolor{urlblue}{Visualization 1}) and perform high resolution 3D reconstruction with tracking results (see \textcolor{urlblue}{Visualization 2}) using our CMFM technique.

\section{Methods}
\label{section3}
\subsection{Image formation model}
The image formation of our CMFM fluorescence system can be modeled as the axial integral of the 2D convolution of the PSF with its corresponding depth layer. In the noiseless case, such model can be written as:
\begin{equation}
g(x,y) = \int_{-\infty}^{\infty}o_z(x,y)*h_z(x,y)dz,
\label{image formation model}
\end{equation}
where $g$ is the observed image, $o$ is the 3D object we want to recover, and $h$ is a z-stack 3D PSF. $o_z$ and $h_z$ are 2D slices of $o$ and $h$ at axial depth $z$, respectively.

Here, we discretize $o$ into $N_x \times N_y \times N_z$ voxels in three dimensions. Each voxel has size $\Delta_x \times \Delta_y \times \Delta_z$. The 2D image $g$ on the detector is sampled with $M_x \times M_y$ pixels. Each pixel has size $\Delta'_x \times \Delta'_y$. For a band-limited system, the Nyquist sampling rate has to be satisfied in order to avoid aliasing. In MFM, the lateral cut-off frequency is $f_c=2NA/\lambda$, and therefore $\Delta'_x  \leq 1/(2f_c)$ and $\Delta'_y  \leq 1/(2f_c)$. For simplicity, we set $\Delta_x = \Delta'_x$ and $\Delta_y = \Delta'_y$ in practice.
In order to match $o$ and $g$ dimensions, $h$ is therefore of $M_x \times M_y \times N_z$ voxels, with each voxel size of $\Delta'_x \times \Delta'_y \times \Delta_z$ .  Then the discrete form of Eq. \eqref{image formation model} can be written as a matrix-vector multiplication form:
\begin{equation}
\mathbf{g} = [\mathbf{H}_1, \mathbf{\cdots}, \mathbf{H}_{N_z}] [\mathbf{o}_1, \mathbf{\cdots}, \mathbf{o}_{N_z}]^{T} = \mathbf{H}\mathbf{o}
\end{equation}
where $\mathbf{g}$ is an $M \times 1$ column vector, in which $M = M_x \times M_y$, $\mathbf{o}$ is an $N \times 1$ column vector, in which $N = N_x \times N_y \times N_z$, and $\mathbf{H}$ is the sensing matrix of a size $M \times N$. Note that each $\mathbf{H}_z$ is a Toeplitz matrix representing 2D convolution, and can be  constructed from $h_z$.

Note that, in our case, two types of noise are considered. First, our imaging process has Poisson noise due to the limited photon budget resulted from (1) the narrower bandpass filter, (2) the lower diffraction efficiency of the MFG and (3) splitting of light. Second, the background noise (room stray light, or the sample itself) is also considered. We assume a uniform background photon noise across all the pixels in the acquired MFM image. Therefore, the image formation model under the Poisson noise and additive background noise model can be expressed as
\begin{equation}
\mathbf{g} =\mathcal{P}\left\lbrace \mathbf{H}\mathbf{o} + \mathbf{b} \right\rbrace ,
\label{Forward model in matrix form}
\end{equation}
where $\mathcal{P}$ represents Poisson statistics originated from signal photons, and $\mathbf{b}$ models the uniform background noise. Note that $\mathbf{b}$ is an $M \times 1$ column vector and each entry of $\mathbf{b}$ is a same constant, denoted by $b$.

\subsection{Joint regularized RL deconvolution}
\label{section3: joint RL-TV algorithm}
Equation (\ref{Forward model in matrix form}) leads to the following likelihood function according to Poisson distribution
\begin{equation}
p(\mathbf{g}|\mathbf{o}, \mathbf{b}) =\prod_{i=1}^{M}\frac{{\left( \mathbf{H}\mathbf{o} + \mathbf{b} \right)_i^{\mathbf{g}_i}}e^{-\left( \mathbf{H}\mathbf{o} + \mathbf{b} \right)_i}}{\mathbf{g}_i!},
\label{Poisson distribution model}
\end{equation}
where $i$ stands for the pixel coordinate in $\mathbf{g}$. In RL deconvolution, the optimal solution $\mathbf{o}^*$ is found from the observation $\mathbf{g}$ by maximizing Eq. \eqref{Poisson distribution model}, or equivalently minimizing its negative logarithm, subject to all the voxels of the restored image have non-negative values, that is,
\begin{equation}
\label{r-l algorithm}
\mathbf{o}^*  =  \arg\;\max_{\mathbf{o}\geq\mathbf{0}}p(\mathbf{g}|\mathbf{o}, \mathbf{b})  = \arg\;\min_{\mathbf{o}\geq\mathbf{0}}f(\mathbf{o}, \mathbf{b}),
\end{equation}
where 
\begin{equation}
f(\mathbf{o}, \mathbf{b}) = \sum_{i=1}^{M} \left[\left( \mathbf{H}\mathbf{o} + \mathbf{b} \right)_i- \mathbf{g}_i\log{\left( \mathbf{H}\mathbf{o} + \mathbf{b} \right)_i}\right]
\end{equation}
is the negative Poisson log-likelihood of $p(\mathbf{g}|\mathbf{o}, \mathbf{b})$ in Eq. \eqref{Poisson distribution model}, in which a constant term $\log{\left(\mathbf{g}_i!\right)}$ is omitted. 

In regularized RL algorithm, a regularizer term is added to the objective function in Eq. \eqref{r-l algorithm}. Here, we consider total variation (TV) regularization because it can avoid the noise amplification problem in RL deconvolution by allowing to recover a smooth and stable solution with sharp edges. The objective function in the TV regularized RL algorithm is therefore written as 
\begin{equation}
\Phi(\mathbf{o},\mathbf{b},\lambda) = f(\mathbf{o}, \mathbf{b}) + \lambda \text{TV}\left(\mathbf{o}\right),
\label{regularized R-L objective}
\end{equation}
where $\lambda$ is the regularization parameter and $\text{TV}(\mathbf{o}) = \sum_{j=1}^N{\lvert\mathbf{\nabla o}_j\rvert}$, in which $j$ denotes the voxel coordinate in the $\mathbf{o}$. Our joint algorithm estimates the 3D image, the background noise and the regularization parameter simultaneously by minimizing $\Phi$ in Eq. \eqref{regularized R-L objective} subject to all the voxels of the restored 3D image have non-negative values, that is,
\begin{equation}
\{ \mathbf{o}^*,\mathbf{b}^*,\lambda^* \} = \arg\; \min_{\mathbf{o}\geq\mathbf{0}, \mathbf{b},\lambda}\Phi(\mathbf{o},\mathbf{b},\lambda).
\label{regularized R-L objective1}
\end{equation}

The minimization of $\Phi(\mathbf{o},\mathbf{b},\lambda)$ in Eq. \eqref{regularized R-L objective1} with respect to all three unknown variables can be performed by an iterative alternating gradient descent method. More specifically, at each iteration, the three variables are updated sequentially, and when one variable is updated, the other two variables are fixed as constants. 

\subsubsection{Estimation of 3D image}
To update $\mathbf{o}$ for $(k+1)$-th iteration, we substitute $\mathbf{o}^k,\mathbf{b}^k$ and $\lambda^k$ estimated from $k$-th iteration in Eq. \eqref{regularized R-L objective}, and take partial derivatives with respect to each voxel $\mathbf{o}^k_j$ in $\mathbf{o}^k$
\begin{equation}
\frac{\partial \Phi(\mathbf{o}^k;\mathbf{b}^k,\lambda^k)}{\partial \mathbf{o}^k_j} = \sum_{i=1}^{M}h_{i,j}-\left[\mathbf{H}^T\frac{\mathbf{g}}{\mathbf{H}\mathbf{o}^k + \mathbf{b}^k}\right]_j - \lambda^k div \left(\frac{\mathbf{\nabla o}^k_j}{\lvert  \mathbf{\nabla o}^k_j\rvert} \right),
\label{partial derivative with respective to o}
\end{equation}
where the divisions are element wise, $h_{i,j}$ is the entry in the $i$-th row and the $j$-th column  of $\mathbf{H}$, and $\sum_{i=1}^{M}h_{i,j} = 1$ if $\mathbf{H}$ is column normalized (which is equivalent to normalizing PSFs $h_z$), and $div$ stands for the divergence operator. Given the partial derivative in Eq. \eqref{partial derivative with respective to o}, We update $\mathbf{o}$ using the 
gradient-based algorithm (or equivalently by using EM
algorithm), defined by \cite{GREEN:90, Dey:06}
\begin{equation}
\begin{aligned}
&\mathbf{o}^{k+1}_j = \left[\mathbf{H}^T\frac{\mathbf{g}}{\mathbf{H}\mathbf{o}^k + \mathbf{b}^k}\right]_j  \frac{\mathbf{o}^k_j}{1-\lambda^k div \left(\frac{\mathbf{\nabla o}^k_j}{\lvert  \mathbf{\nabla o}^k_j\rvert} \right)}, \\
& \qquad \qquad \qquad \qquad  \mathbf{o}^{k+1}_j \geq 0, 
\label{update rule for o}
\end{aligned}
\end{equation}
where the second line in Eq. \eqref{update rule for o} is to enforce the non-negativity on $j$-th reconstructed voxel. Note that for $1\leq j \leq N$, all the voxels in $\mathbf{o}$ can be updated simultaneously by storing them in a vector. 

\subsubsection{Estimation of the uniform background}
When we have $\mathbf{o}^{k+1}$, we substitute $\mathbf{o}^{k+1},\mathbf{b}^k$ and $\lambda^k$  in Eq. \eqref{regularized R-L objective}, and take partial derivatives with respect to $b$
\begin{equation}
\frac{\partial \Phi(\mathbf{b}^k; \mathbf{o}^{k+1},\lambda^k)}{\partial b^k} = \sum_{i=1}^{M} \left\{ 1-\left[\frac{\mathbf{g}}{\mathbf{H}\mathbf{o}^{k+1} + \mathbf{b}^k}\right]_i \right \}.
\label{partial derivative with respective to b}
\end{equation}
Note that each pixel in the captured MFM image is assumed to have the same background value $b$. Then we update $b$ using the same gradient-based algorithm as is in 3D image estimation
\begin{equation}
b^{k+1} = \left\{ \frac{1}{M}\sum_{i=1}^{M}\left[\frac{\mathbf{g}}{\mathbf{H}\mathbf{o}^{k+1} + \mathbf{b}^k}\right]_i\right \}b^k.
\label{update rule for b}
\end{equation}

\subsubsection{Estimation of optimal regularization parameter}

\begin{algorithm}[t!]
\caption{Joint regularized RL algorithm for MFM }\label{Joint regularized R-L deconvolution1}
\begin{algorithmic}[1]
\State \textbf{inputs:} $\mathbf{g}$ and $\mathbf{H}$ \Comment{2D MFM data and 3D PSFs}
\State \textbf{initialize:} $\mathbf{o}^0,\mathbf{b}^0,$ and $\lambda^0$
\State set $k=0$
\While{not convergence}
\State update $\mathbf{o}^{k+1}$ using $\mathbf{o}^k,\mathbf{b}^k$ and $\lambda^k$ in Eq. \eqref{update rule for o}
\State update $\mathbf{b}^{k+1}$ using $\mathbf{o}^{k+1},$ and $\mathbf{b}^k$ in Eq. \eqref{update rule for b}
\State update $\lambda^{k+1}$ using $\mathbf{o}^{k+1},$ and $\mathbf{b}^{k+1}$  in Eq. \eqref{update rule for l2}
\State $k = k+1$
\EndWhile\label{euclidendwhile}
\State \textbf{return:} $\mathbf{o}^*,\mathbf{b}^*,$ and $\lambda^*$ \Comment{joint estimations of 3 unknowns}
\end{algorithmic}
\end{algorithm}
The approach of estimating optimal $\lambda$ is based on the fact that when the optimal solution $\mathbf{o}^*$ is found, 
the partial derivatives with respect to each restored voxel $\mathbf{o}^k_j$ in Eq. \eqref{partial derivative with respective to o} should be equal or close to zero. So the $\lambda$ is obtained by minimizing the sum of the square of Eq. \eqref{partial derivative with respective to o} over all the reconstructed voxels \cite{Laasmaa:11} 
\begin{equation}
\lambda^{k+1} = \arg \; \min_{\lambda} \sum_{j=1}^N \Vert \frac{\partial \Phi(\mathbf{o}^{k+1};\mathbf{b}^{k+1},\lambda)}{\partial \mathbf{o}^{k+1}_j} \Vert^2.
\label{minimize partial derivatives}
\end{equation}
Because the objective function in Eq. \eqref{minimize partial derivatives} is a quadratic form of $\lambda$, the close form solution of $\lambda$ can be found by equating the derivatives of Eq. \eqref{minimize partial derivatives} with respect to $\lambda$ to zero
\begin{equation}
\lambda^{k+1} = \frac{\sum_{j=1}^N \left\{1- \left[\mathbf{H}^T\frac{\mathbf{g}}{\mathbf{H}\mathbf{o}^{k+1} + \mathbf{b}^{k+1}}\right]_j \right \}div \left(\frac{\mathbf{\nabla o}^{k+1}_j}{\lvert  \mathbf{\nabla o}^{k+1}_j\rvert} \right) }{\sum_{j=1}^N {\left[ div \left(\frac{\mathbf{\nabla o}^{k+1}_j}{\lvert  \mathbf{\nabla o}^{k+1}_j\rvert} \right) \right]}^2}.
\label{update rule for l2}
\end{equation}
The joint regularized RL deconvolution algorithm is summarized in Algorithm \ref{Joint regularized R-L deconvolution1}. 
We performed a series of simulations to evaluate the effectiveness and convergence of the joint regularized RL algorithm for our CMFM system. The simulation results and convergence plots are shown in Appendix A.

\section{System analysis}
\label{system analysis}
A schematic of our CMFM imaging setup is shown in Fig. \ref{fig:overview}(a). A regular microscope is augmented by a $4f$ relay system (L1 = 200mm and L2 = 400mm). The total magnification of the whole system is $120\times$ with the use of $60\times$ objective lens.  An electron multiplying charge coupled device (EMCCD) has $1024 \times 1024$ pixels with pixel pitch $13\text{um} \times 13\text{um}$. Thus, each pixel corresponds to a size of $108\text{nm} \times 108\text{nm}$ in the object domain, which satisfies the Nyquist sampling criterion. A multifocal gratings (MFG) is placed at the Fourier plane of the $4f$ relay system. As an example, schematics of the MFG used in our first experiment is shown in Fig. \ref{fig:overview}(b). This MFG is optimized and designed to produce $3\times3$ differently focused tile images on a single 2D detector as shown in Fig. \ref{fig:overview}(c), with an equal focal step $\Delta z$ of $0.25\text{um}$ between adjacent tiles and almost even light energy distribution of $[7.56, 7.48, 7.21, 7.47, 7.62, 7.47, 7.21, 7.48, 7.56]$ percent for each tile. A total diffraction efficiency of the MFG is therefore $67.06\%$, which is close to the theoretical maximum efficiency.  To avoid the overlapping between tile images, a field stop slightly smaller than the tile width is placed at the intermediate image plane to reduce the lateral FOV [see Fig. \ref{fig:overview}(a)]. 

Next, we provide analysis on the Point Spread Function (PSF), the chromatic aberration (CA), and the FOV of our CMFM system. 

\subsection{Point Spread Function (PSF) and resolution}
\label{PSF and resolution}
\begin{figure}[t!]
\centering
  \includegraphics[width=0.8\linewidth]{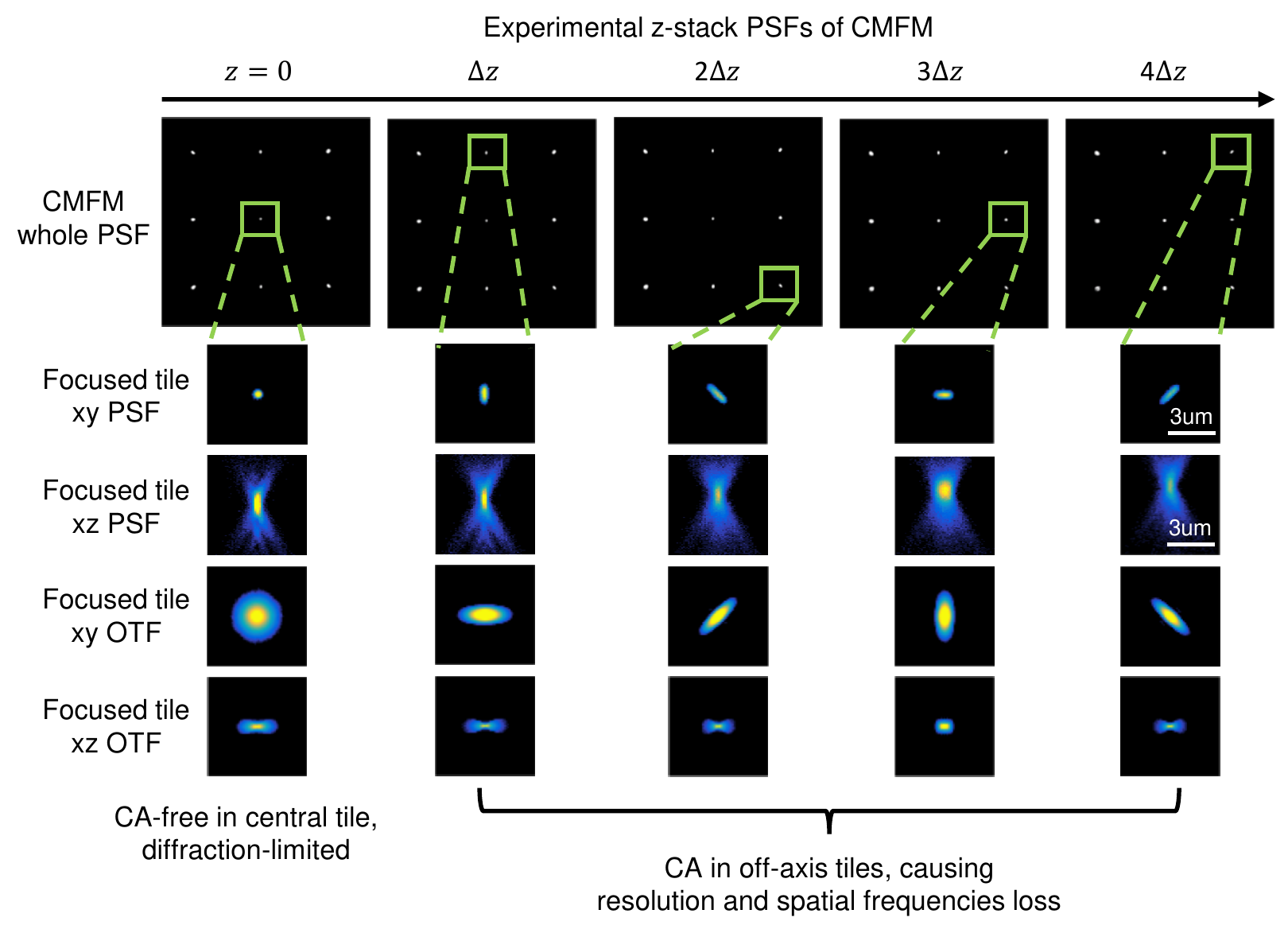}
  \caption{A z-stack 3D PSFs of CMFM, measured from a $170$nm fluorescent bead. Top row: CMFM lateral PSFs imaged under five different axial positions (columns). each PSF consists of one focused image (outlined by a green box) and eight out-of-focus version images of the bead.  xy and xz PSFs (second and third rows) and corresponding OTFs (bottom two rows) of five differently focused tiles (columns). The focal shift property of CMFM can be observed from xz PSFs (third row), verifying that CMFM is capable of capturing a focal stack instantaneously. Although the central tile's PSF CA-free (first column), the off-axis tiles' PSFs suffer from directional CA (second to last columns) due to geometry. The lateral spatial frequencies that are lost by CA are shown in xy OTFs (fourth row).}
  \label{MFM tile PSF}
\end{figure}

We image a z-stack of a small fluorescent bead (about $170$nm in diameter), which approximates an ideal light-emitting point, to measure PSF. Figure \ref{MFM tile PSF} (first row) shows CMFM lateral PSF imaged under five different axial positions (columns). Each lateral PSF consists of $3\times3$ tiles, with one tile in focus (outlined by a green box; also zoomed in second row) and eight tiles out of focus. From left to right, as the focus depth increases, we can observe a focal shift on the axial direction of the x-z plots of PSFs (third row). The Optical Transfer Functions (OTFs) are plotted as reference (bottom two rows).

We measure the full-width half-magnitude (FWHM) values to quantify the resolution. Note that the PSFs are affected by chromatic aberration and the lateral resolution is anisotropic. Therefore, we perform a principle component analysis (PCA) of xy PSFs. The FWHM values of those five tiles' PSFs (second row) are $\{0.32, 0.87, 1.17, 0.87, 1.17\}\text{um}$ in the CA direction, and $[0.32, 0.37, 0.38, 0.37, 0.36]\text{um}$ in the perpendicular direction. The axial FWHM values are $\{1.03, 1.39, 1.50, 1.39, 1.55\}\text{um}$. The variation of the in-focus PSF sizes results from directional chromatic aberration. Based on the PCA results, each horizontal and vertical diffraction order tile has a resolution loss of $0.55\text{um}$ due to CA. Each diagonal diffraction order tile has a resolution loss of $0.85\text{um}$ due to geometry. We show, in the next sub-section, that this observation is consistent with the theoretical calculation. 

\subsection{Chromatic aberration (CA)}
\label{CA}
Due to the diffractive nature of the MFG, the CMFM system suffers from obvious CA effect, which causes the loss of the intensity and resolution. Mathematically, the lateral CA per tile can be expressed as \cite{Kuan:18}:
\begin{equation}
\delta_x = \frac{mf\Delta \lambda}{d_x}; \delta_y = \frac{nf\Delta \lambda}{d_y},
\label{lateral CA}
\end{equation}
where $m$ and $n$ are horizontal and vertical diffraction orders of the tile, $f$ is the focal length of the second lens of the $4f$ system, $d_x$ and $d_y$ are periods of the MFG in the $x$ and $y$ directions, and $\Delta \lambda$ is the bandpass spectrum of the emission collected by the detector. The axial CA for each tile, i.e., different wavelengths are focused at different distances from the lens, can also be expressed as \cite{Kuan:18}:
\begin{equation}
\delta_z = \frac{\Delta \lambda}{\lambda_c}f_{m,n},
\label{axial CA}
\end{equation}
where $\lambda_c$ the wavelength used to design the MFG, and $f_{m,n}=(m+ln)\Delta z$ is the focusing distance of each individual tile. 

From Eq. \eqref{lateral CA} and Eq. \eqref{axial CA}, we can see that for the central tile where $m=n=0$, $\delta_x=\delta_y=\delta_z=0$, which means the central tile's PSF and image are free of CA. However, for off-axis tile, both lateral and axial CA exist. Each horizontal and vertical diffraction oder tile has a dispersion of $mf\Delta \lambda /d_x$, which is equal to $0.59\text{um}$ when $m=1$ for our CMFM system of $f=400\text{mm}$, $\Delta \lambda = 10$nm, $d_x = d_y =56$um and $\hat{M}=120$. Each diagonal diffraction order tile has a dispersion of $\sqrt{m^2+n^2}f\Delta \lambda /d_x$. In addition, the dispersion direction varies for different diffraction order tiles due to geometry. 
If left uncorrected, the CA would cause a loss of the the intensity and resolution in the chromatic dispersion direction. In principle, CA can be optically corrected by using CA corrective optics. However, the corrective optics increases system cost and complexity\cite{Abrahamsson:13}. Here, we show that CA can be effectively compensated by \emph{computation}. This is possible because the CA is directional due to geometry of the tile distribution in our CMFM system. For example, in Fig. \ref{MFM tile PSF}, the eight non-centric tiles exhibit directional stretching towards the image center. At depth $z=0$ (first column), the PSF is CA-free. On the other focal planes, from $\Delta z$ to $4\Delta z$ (second column to last column), even though the green boxes indicate in-focus tiles, the PSFs stretch in different directions at different diffraction orders. However, the stretching does not occur at the perpendicular direction of CA. This implies that although a certain tile is affected by CA on one direction, the perpendicular direction is less affected and can be used to compensate another tile which suffers CA at this direction. Therefore, by jointly utilizing all the tiles in the model, 3D deconvolution can preserve 2D spatial frequencies that are lost by CA.

\begin{figure}[t!]
  \includegraphics[width=0.8\linewidth]{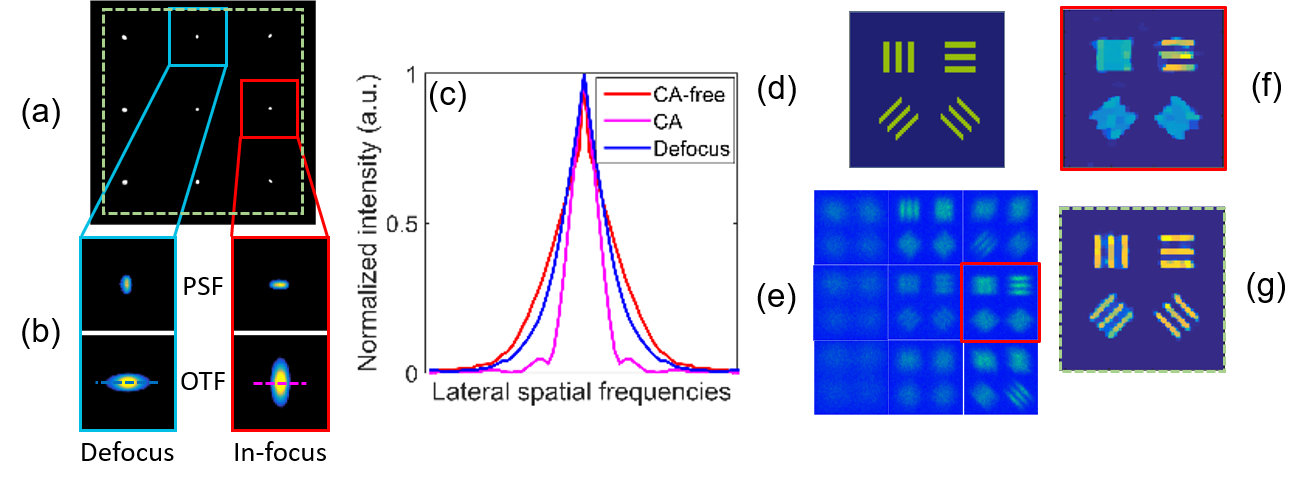}
  \centering
  \caption{CA blur vs defocus blur. An in-focus tile with CA blur (red) and a defocus blur (blue) are highlighted in (a), whose PSFs and OTFs are shown in (b). (c) plots a comparison of linecuts indicated by blue and magenta lines in (b). For reference, a linecut in CA-free central tile's OTF (shown in first column and fourth row of Fig. \ref{MFM tile PSF}) is also plotted (red). A reconstruction comparison is shown in the right panel. (d) Object image. (e) Observation image (for visualization purpose, each tile image is cropped). (f) Reconstruction using only in-focus PSF. (g) Reconstruction using all the PSFs.}
  \label{CA blur vs out-of-focus blur}
\end{figure}
\emph{CA blur vs defocus blur.} Note here that the blur resulted from CA is different than the defocus blur. Fig. \ref{CA blur vs out-of-focus blur} presents a comparison of the two types of blurs. In Fig. \ref{CA blur vs out-of-focus blur}(a), the red box highlights the in-focus tile. This in-focus tile is not located at the image center and has CA blur in horizontal dimension. Meanwhile, the blue box highlights the tile that suffers both defocus blur and CA blur but in vertical dimension. However, in horizontal dimension, this out-of-focus tile only has defocus blur.  An Optical Transfer Function (OTF) comparison is shown in Fig. \ref{CA blur vs out-of-focus blur}(b). The defocus blur size is smaller than the CA blur size in horizontal direction, resulting in a wider OTF lobe, which implies that higher resolution reconstruction can be achieved by using all the information provided from the PSFs. On the right panel, we present numerical simulation results. The simulated resolution target [Fig. \ref{CA blur vs out-of-focus blur}(d)] contains bars pointing four directions. By using only the in-focus PSF, as shown in Fig. \ref{CA blur vs out-of-focus blur}(f), only horizontal bars can be resolved, as the high frequency has only been preserved in vertical direction [Fig. \ref{CA blur vs out-of-focus blur}(b)]. On the other hand, by making full use of the PSFs, signals on different directions can be well-recovered [Fig. \ref{CA blur vs out-of-focus blur}(g)]. 

\begin{figure}[htbp]
\begin{center}
  \includegraphics[width=0.68\linewidth]{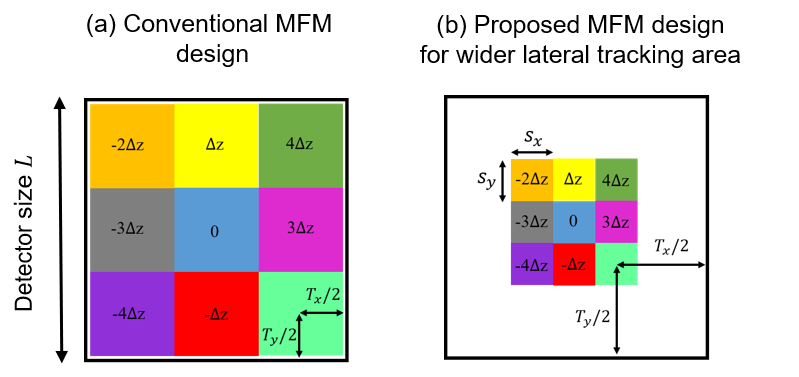}
  \centering
  \caption{(a) Conventionally, a MFG is designed to project the diffraction pattern occupying the entire imaging space. (b) We propose to use a smaller tile spacing, so as to achieve a small lateral FOV that can be tracked over a large area for MFM tracking applications. }
  \label{fig:relaxing lateral FOV}
\end{center}
\end{figure}
\subsection{Field-of-View (FOV)}
\label{FOV}
In conventional MFM designs [Fig. \ref{fig:relaxing lateral FOV}(a)], the lateral FOV is the same as the lateral size of each tile image, which can be expressed in the sample domain as:
\begin{equation}
FOV_{x} = \frac{L_x}{l\hat{M}}; FOV_{y} = \frac{L_y}{l\hat{M}},
\label{lateral FOV}
\end{equation}
where $L_x$ and $L_y$ are detector size in the $x$ and $y$ directions, $l$ is the number of tiles in each dimension, and $\hat{M}$ is the magnification of the MFM. The axial FOV is limited by the focused imaging range of the designed MFG as
\begin{equation}
FOV_z = (l^2-1)\Delta z.
\label{axial FOV}
\end{equation}
From Eq. \eqref{lateral FOV} and Eq. \eqref{axial FOV}, we can see that by increasing the number of tiles $l^2$, the axial FOV will increase but at the expense of the reduced lateral FOV for a fixed detector size. Thus, MFM trades its lateral FOV for the depth resolution. The maximum recovered 3D volume or tracking space in MFM is $FOV_{x} \times FOV_{y}\times FOV_{z}$.

\subsubsection{Enlarging lateral tracking space}
In conventional MFM designs, large tile spacings are used to ensure that large objects can be imaged [see Fig. \ref{fig:relaxing lateral FOV}(a)]. However, this comes at the cost of a large reduction in lateral tracking area. This, however, is not optimal for the MFM 3D tracking applications where the objects of interest (bacteria or molecule, etc.) are quite small but move freely in a large 3D space. In single molecule tracking, larger tracking space is more desirable than the lateral FOV.

By modifying our MFG design to produce a smaller tile spacing, we can achieve a small lateral FOV that can be tracked over a large area [see Fig. \ref{fig:relaxing lateral FOV}(b)]. The comparison of MFM image diagrams obtained by two different design methods are shown in Fig. \ref{fig:relaxing lateral FOV}. When optimizing the MFG for tracking applications, the dimensions of the trackable area
\begin{equation}
T_{x} = \frac{L_x - (l-1) s_x}{\hat{M}}; T_{y} = \frac{L_y - (l-1) s_y}{\hat{M}},
\label{lateral tracking space}
\end{equation}
in $x$ and $y$ dimensions, where $s_x$ and $s_y$ are tile spacing in x and y dimensions. Note that  Eq. \eqref{lateral tracking space} is a general form of  Eq. \eqref{lateral FOV}. From Eq. \eqref{lateral tracking space}, we can clearly see that the lateral tracking area $T_x$ and $T_y$ can be enlarged by designing a smaller tile spacing $s_x$ and $s_y$. Therefore, our method can achieve a larger lateral tracking area for MFM tracking applications.

\subsubsection{Numerical validation}
\label{section: larger FOV CMFM}
\begin{figure}[hbtp]
\begin{center}
  \includegraphics[width=0.68\linewidth]{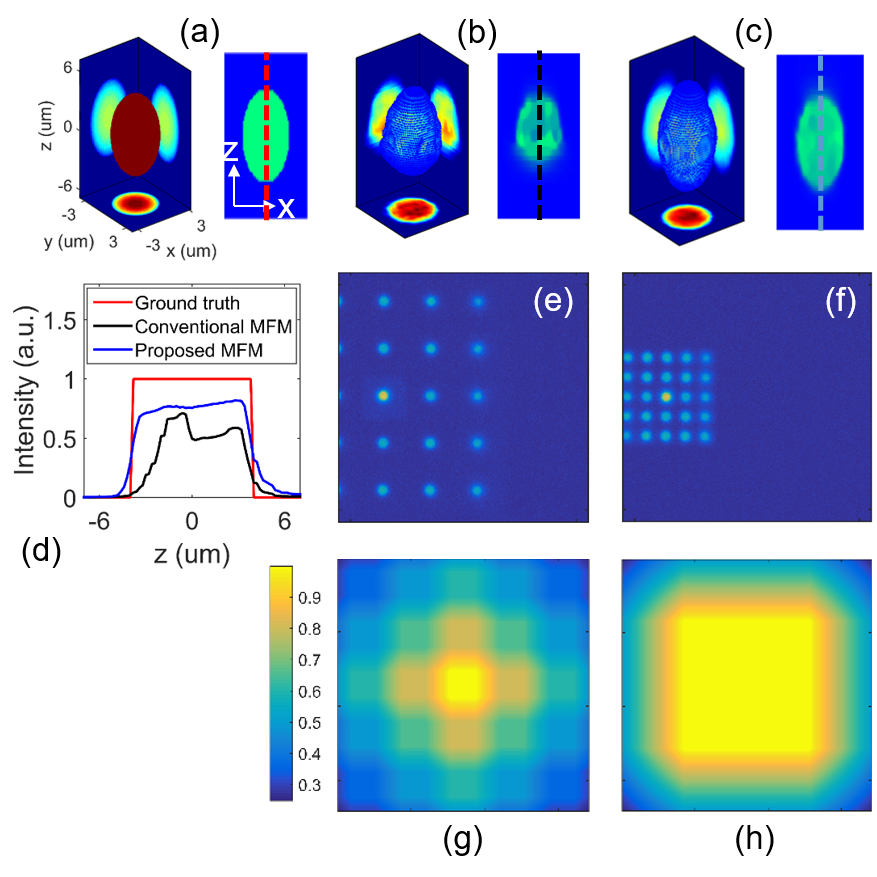}
  \centering
  \caption{Simulations showing the capability of the proposed MFM of achieving larger lateral tracking space than conventional MFM does. (a) The synthetic 3D ground truth of an ellipsoid (left) and its $xz$ slice (right). The center of the ellipsoid is $35.8$um away from the center of the detector in $x$ direction. MFM measurements (e-f) and corresponding reconstructions (b-c) by different design methods. (d) 1D axial profile comparison between ground truth (red) and reconstructions (black and blue). It is clear that the ellipsoid is reconstructed poorly from conventional MFM method while our design provides a good reconstruction. Signal loss as a function of lateral position of the tracked object for conventional (g) and our designed MFM (h). Similar to vignetting effect, the signal falls off when approaching the edges. Our proposed design alleviates peripheral signal loss and achieves an enlarged lateral tracking area.}
  \label{fig: larger FOV MFM}
\end{center}
\end{figure}

We simulated a 3D tracking space of $1024\times1024\times71$ voxels with each voxel size of $108\text{nm} \times 108\text{nm} \times 200\text{nm}$. The scene consists of an ellipsoid [Fig. \ref{fig: larger FOV MFM}(a)] with a diameter of $4.32$um in $x$ and $y$ directions and $8$um in $z$ direction, which is similar to the size of a bacterium in our experiment [see Fig. \ref{experiment results}]. The $xz$ slice and 1D axial profile of the ellipsoid are shown in Fig. \ref{fig: larger FOV MFM}(a) (right) and Fig. \ref{fig: larger FOV MFM}(d) (red line), respectively. The center of the ellipsoid is placed at $35.8$um horizontally from the center of the detector.

For conventional MFM, we used an experimentally captured z-stack PSF for the simulation. The PSF consists of $5\times5$ tiles with a tile spacing of $205$ pixels in both $x$ and $y$ dimensions.  The 2D MFM measurement [Fig. \ref{fig: larger FOV MFM}(e)] was generated based on the forward model of Eq. \eqref{Forward model in matrix form}. The maximum signal and background photon counts are set to be $100$ and $5$, respectively. The Poisson noise was then added to the measurement by using Matlab's Poisson random number generator. For the proposed MFM, the PSF was generated by cropping the central $101\times101$ pixels region of each tile of the experimentally acquired MFM PSF and then recombining the cropped small tiles. As a result, the tile spacing is $101$ pixels instead of $205$ pixels. The measurement [Fig. \ref{fig: larger FOV MFM}(f)] of this new MFM design was also generated based on Eq. \eqref{Forward model in matrix form} and then corrupted by background and Poisson noise.

In conventional MFM, since the horizontal position of the synthetic ellipsoid exceeds the tile spacing, the left two-column tiles produced by MFM which contains axial depth information from $-12\Delta z$ to $-3\Delta z$ are shifted out of the detector's FOV and therefore can not recorded by the detector [Fig. \ref{fig: larger FOV MFM}(e)]. However, in our new MFM design, the $5\times5$ tiles  are recorded in their entirety by the detector [Fig. \ref{fig: larger FOV MFM}(f)]. For reconstruction, we recovered a 3D space on a grid of $1024\times1024\times71$ with each grid size being $108\text{nm} \times 108\text{nm} \times 200\text{nm}$ for both methods.  The reconstructed ellipsoid (cropped from the larger 3D recovery space) and its $xz$ slice by different design methods are shown in Figs. \ref{fig: larger FOV MFM}(b) and (c), respectively. The 1D axial profiles of the reconstructed ellipsoids are also compared in Fig. \ref{fig: larger FOV MFM}(d) (black and blue lines). We can clearly see that since the left two-column tiles of the ellipsoid are missing in the conventional MFM measurement, the ellipsoid is reconstructed poorly while our design provides a good reconstruction. We also compare the signal loss as a function of lateral position of the tracked object in Figs. \ref{fig: larger FOV MFM}(g) and (h). Similar to vignetting effect, the signal falls off when approaching the edges. Our proposed design alleviates peripheral signal loss and preserves full-sized PSF array over an enlarged lateral space.

\section{Experimental results}
\label{experiment}
\begin{figure}[hbtp]
\includegraphics[width=0.65\linewidth]{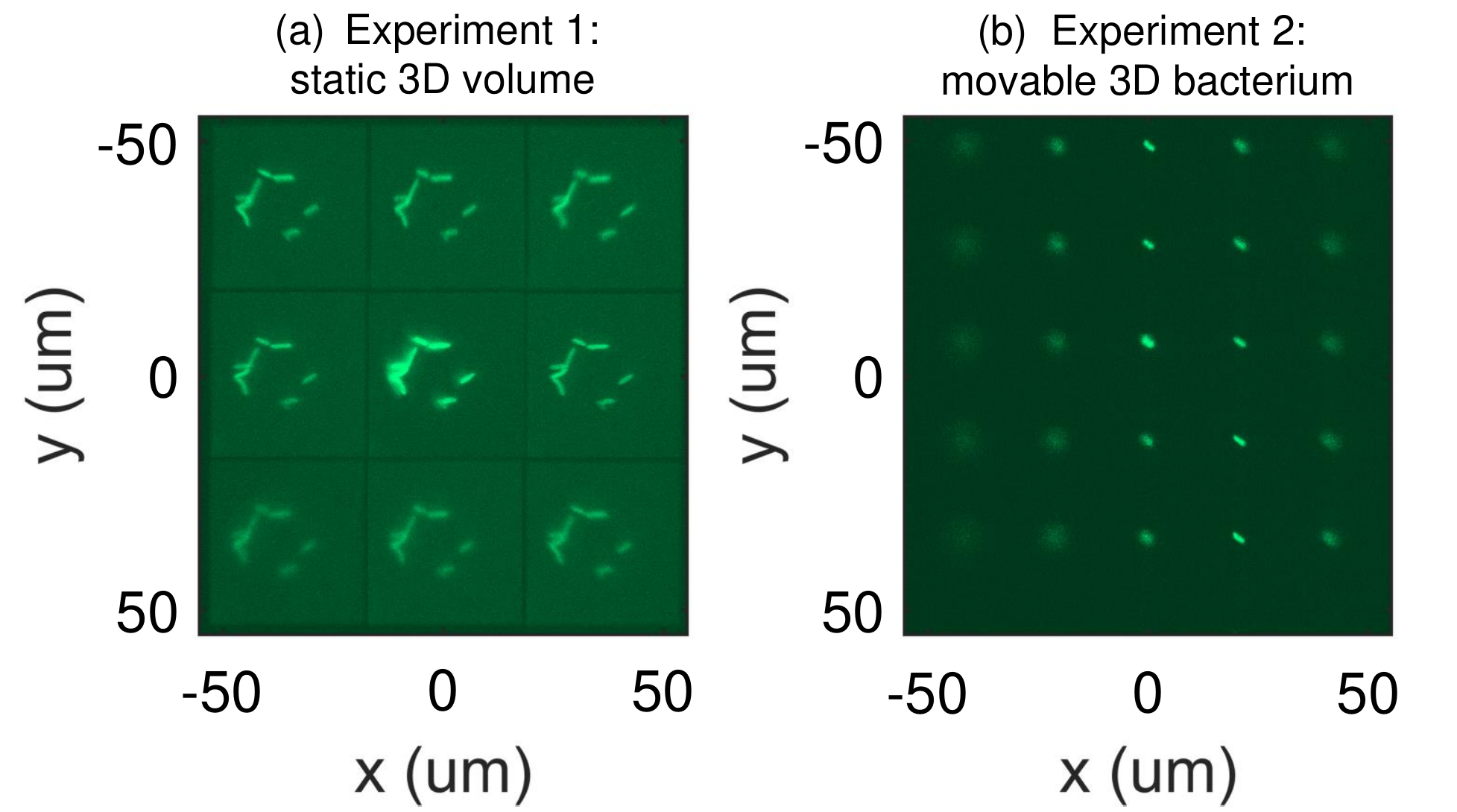}
\centering
  \caption{Two experimental snapshot MFM raw images. (a) Experiment 1: snapshot captured 2D MFM image of multiple static periplasms by using an MFG with $9$ focal planes under exposure time of $0.5$s. (b) Experiment 2: a frame from an MFM video of a moving bacterium captured at $25$ fps by using an MFG with $25$ focal planes. The raw MFM video is shown in \textcolor{urlblue}{Visualization 1}.}
  \label{Experimental snapshot MFM images}
\end{figure}

We present two experimental results with different biological samples to test the spatial and temporal performance of our system. As shown in Fig. \ref{Experimental snapshot MFM images}, we divide the whole image into different sets of tiles by modifying the MFG pattern. However, the focal step $\Delta z$ is designed to be $0.25$um for both cases. In Fig. \ref{Experimental snapshot MFM images}(a), the image is divided into $3\times 3$ tiles so that 9 focal planes are in focus, while in Fig. \ref{Experimental snapshot MFM images}(b), $5\times 5$ focal planes are obtained for the purpose of tracking. 

\subsection{Static scene: CMFM \emph{vs} confocal microscopy}
\label{Static scene}
\begin{figure}[ht!]
  \includegraphics[width=0.75\linewidth]{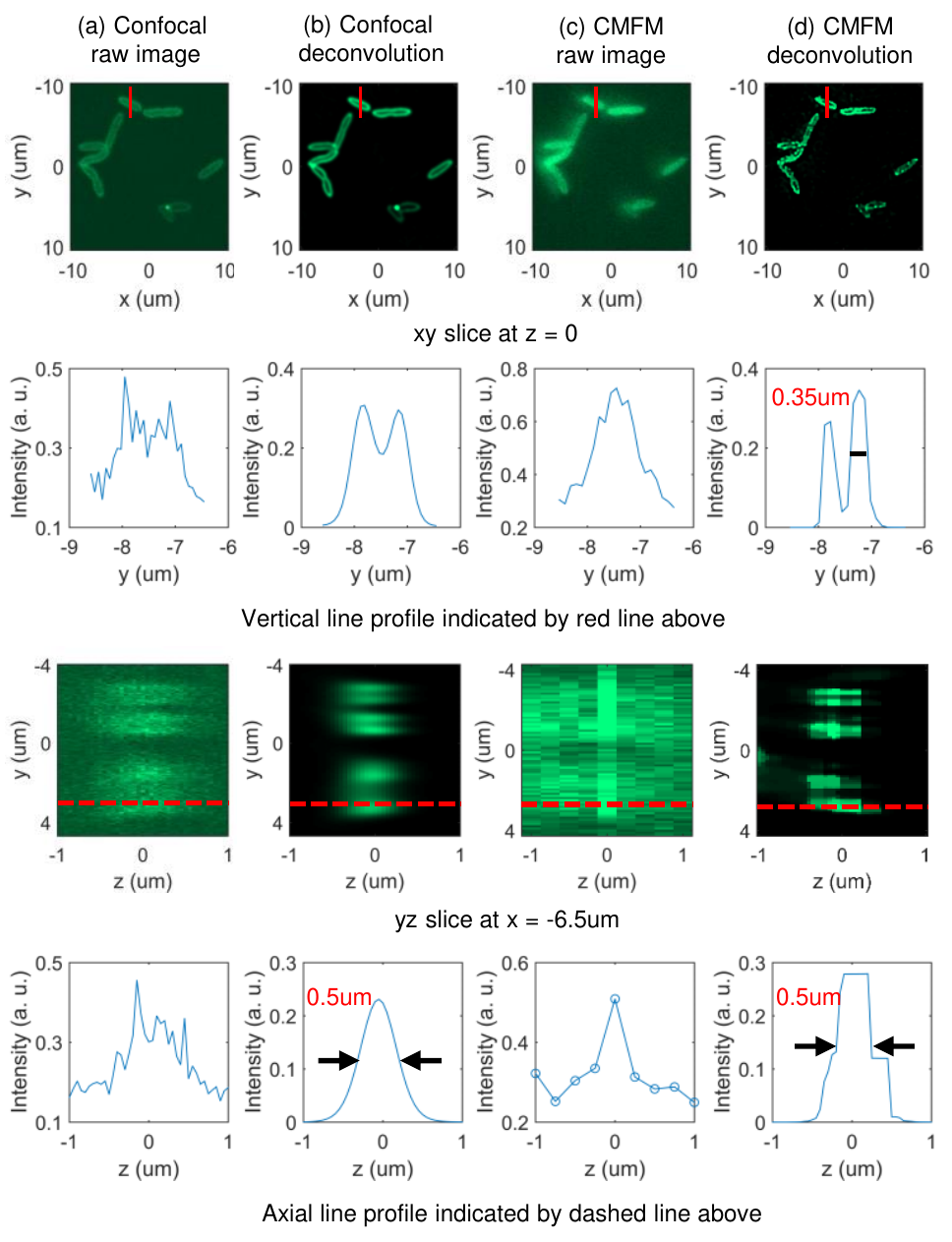}
  \centering
  \caption{Proposed computational 3D reconstruction of CMFM image in comparison with confocal deconvolution results. (a) Confocal raw data and (b) its deconvolution results. (c) CMFM raw data and (d) its computational reconstruction results. In (a), confocal scan is taken with a dual spinning disk confocal microscope (Model: CSU-W1) with the total acquisition time of $20$s, while in (c), the CMFM raw data is captured in a single exposure of $0.5$s. The lateral resolution of the proposed computational reconstruction is about $0.35$um (second row of d) and the axial resolution is about $0.5$um (fourth row of d), which are comparable with those achievable with confocal deconvolution microscopy (second and fourth rows of b).}
  \label{periplasm reconstruction}
\end{figure}
We prepare several periplasms (static) as our imaging sample for 3D spatial resolution characterization. The captured raw image is shown in Fig. \ref{Experimental snapshot MFM images}(a). The exposure time is $0.5$s.  The 3D PSFs are measured by recording a z-stack of a small fluorescent bead with z focal step of $50$nm. The 3D image are reconstructed on a $300\times300\times51$ grid with each grid size being $108\text{nm} \times 108\text{nm} \times 50\text{nm}$. Figure \ref{periplasm reconstruction} shows our computational 3D reconstruction [Fig. \ref{periplasm reconstruction}(d)] in comparison with the captured focal stack [Fig. \ref{periplasm reconstruction}(c)] assembled by z-stacking $9$ tiles from the 2D raw MFM image. For ground truth, we captured the 3D image of periplasms using a scanning confocal microscope [Fig. \ref{periplasm reconstruction}(a)] and performed deconvolution of the confocal image [Fig. \ref{periplasm reconstruction}(b)] by using a commercial Huygens software. For each 3D image, a XY slice at $z=0$ (first row) and a YZ slice at $x=-6.5$um (third row) are displayed. The comparisons of vertical line profiles indicated by the red solid line in XY slice, and comparison of axial profiles indicated by the red dotted line in YZ slice are shown in the second and fourth rows, respectively. From Fig. \ref{periplasm reconstruction}(c), we can see that without computational reconstruction, the 3D image assembled by z-stacking nine tile images from 2D MFM measurement is degraded by noise, out-of-focus blur and low spatial resolution in all three dimensions. The outer membranes are blurred and the empty space between them cannot be recognized. In addition, due to the out-of-focus blur, the FWHM of the axial profile is about $2$um. However, after deconvolution, the reconstructed 3D image [Fig. \ref{periplasm reconstruction}(d)] is much cleaner, less out-of-focus blur, and more importantly, has high resolution in three dimensions. The empty space between outer membranes can be clearly recognized. The outer membranes which are vertically blurred to a single peak can be well separated by a dip of $84\%$ between two peaks with the lateral FWHM of $0.35$um [second row of Fig. \ref{periplasm reconstruction}(d)], which is consistent with FWHM of the measured PSF. Due to removal of out-of-focus blur, the axial FWHM is about $0.5$um [fourth row of Fig. \ref{periplasm reconstruction}(d)], with about two times improvement over that of the raw MFM image. Note that we compare $0.5$s captures with our CMFM instrument to a $20$s confocal scan taken with a dual spinning disk confocal microscope (Model: CSU-W1) made by Yokogawa Electric Corporation. Our CMFM results show similar 3D image quality, but achieve a 40x reduction in acquisition time. The automatically recovered background noise and optimal regularizer parameter at each iteration of the joint RL-TV deconvolution process is shown in Appendix B Fig. \ref{fig: background and lambda}. 

\emph{Cautionary remarks.} Note that although both our CMFM and confocal deconvolution microscopy achieve $0.5$um axial resolution in the experiment, the principle of the axial resolution improvement is different for two techniques. Confocal microscopy increases axial resolution by means of using a (or multiple) spatial pinhole(s) to block out-of-focus light in image detection process. Therefore, the axial extent of confocal PSF is narrower than that in the widefield microscope, and thus a high contrast and resolution image can be obtained. However, for CMFM, the axial resolution improvement is based on sparsity-based super-resolution microscopy techniques \cite{hugelier2016sparse,hugelier2017improved,szameit2012sparsity} by utilizing the signal sparsity in the arbitrary known domain (i.e., gradient domain in our case). Recently, a new method called SPARCOM \cite{Solomon:18}: sparsity-based super-resolution correlation microscopy by utilizing sparsity in the correlation domain, is also reported to achieve spatial resolution comparable to PALM and STORM.

\subsection{3D tracking of moving bacterium}

\begin{figure}[htbp]
\begin{center}
  \includegraphics[width=0.8\linewidth]{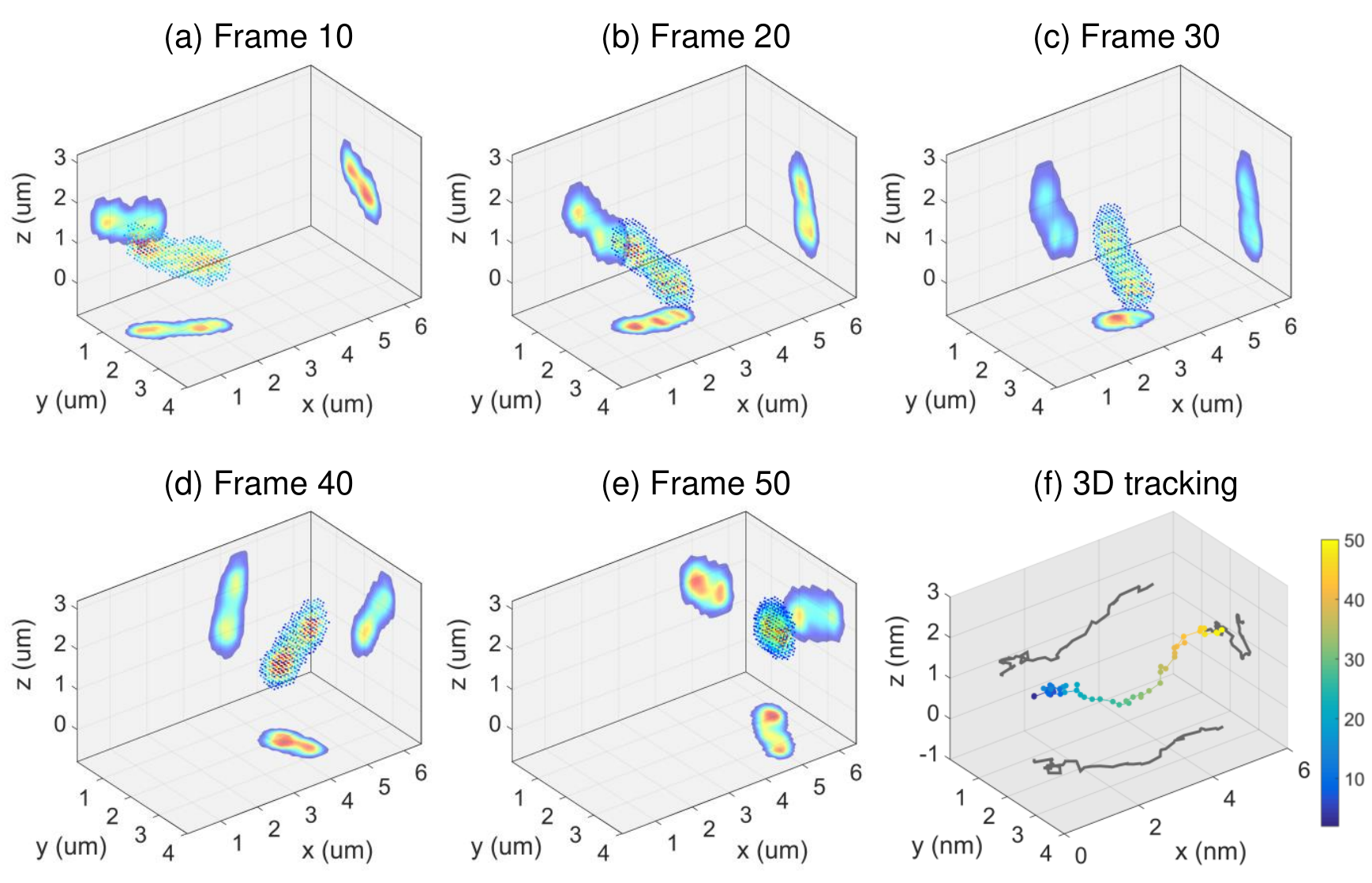}
  \caption{Experimental 3D reconstructions of a movable bacterium.  A raw MFM video (shown in \textcolor{urlblue}{Visualization 1}) was captured at $25$fps as the bacterial moves in 3D space. The computational 3D reconstruction was performed for each video frame. Five out of sixty frames reconstruction is shown in (a-e). (f) 3D trajectory of the bacterium by computing and tracking its center of mass for each frame reconstruction. The colorbar indicate the frame index over time. The complete 3D video reconstruction from the first frame to the last frame is shown in \textcolor{urlblue}{Visualization 2}.}
  \label{experiment results}
\end{center}
\end{figure}

In the second experiment, we demonstrate the 3D video reconstruction of a moving bacterium (see \textcolor{urlblue}{Visualization 2}). The MFG is modified to focus on $5\times 5$ planes, as shown in Fig. \ref{Experimental snapshot MFM images}(b). A video is captured at $25$ frame per second (fps) (see \textcolor{urlblue}{Visualization 1}). According to our first experimental reconstruction where the axial FWHM can achieve $0.5$um, we reconstruct the 3D image of the bacterium on a grid of $150\times150\times41$ with each grid size being $108\text{nm} \times 108\text{nm} \times 200\text{nm}$, which corresponds to a FOV of $16.2\text{um} \times 16.2\text{um} \times 8\text{um}$ in 3D space for each video frame. Figures \ref{experiment results}(a-e) show five out of fifty frames 3D reconstructions. Note that for the visualization purpose, we cropped the reconstructed 3D volume and just showed $4\text{um} \times 6\text{um} \times 4\text{um}$ region around the bacterium. By computing the center of mass for each frame reconstruction, a 3D trajectory of the moving bacterium can be tracked [Fig. \ref{experiment results}(f)]. The colorbar in Fig. \ref{experiment results}(f) indicates the frame index over time. The automatically recovered background value and the optimal regularizer parameter for each video frame is shown in Appendix B Fig. \ref{fig: background and regularizer parameters estimation}.



\section{Conclusion}

We have presented computational multi-focus microscopy (CMFM), a framework that balances the imaging system design and computational processing to achieve single-shot 3D microscopy. Our optical design significantly reduces the system complexity and experimental alignment by discarding CA correction optics. We correct for CA computationally rather than optically. This comes from the fact that CA occurs in different directions at different diffraction order tiles. Therefore, by jointly utilizing all the tiles in the model, 3D deconvolution can preserve 2D spatial frequencies that are lost by CA.  By incorporating TV regularization, our algorithm can not only compensate for CA, but also perform high quality 3D reconstructions from noisy data. We build on a joint regularized RL deconvolution algorithm and incorporate two types of noise. Notice that our proposed algorithm is free of any parameter tuning by automatically estimating the background noise and the optimal reularization parameter using alternating gradient descent. 

We experimentally demonstrate that the out-of-focus blur along $z$ can be significantly suppressed and the axial resolution as high as $0.5$um is achievable. The lateral resolution of $0.35$um, which is consistent with the diffraction-limited resolution, is also experimentally demonstrated. A high resolution 3D video of a movable bacterium at 25fps is also computationally reconstructed to verify the proposed CMFM framework. 

Finally, we propose a new design method of MFG to enlarge the lateral tracking area of MFM tracking applications without sacrificing its axial FOV and single-shot capture speed. The benefits of the simple system design and high resolution image recovery offered by the proposed CMFM will broaden its applications in 3D single-shot imaging.

\section*{Appendix A: algorithm evaluation}

\label{section: algorithm analysis}
\begin{figure}[t!] 
\begin{center}
\includegraphics[width=0.62\linewidth]{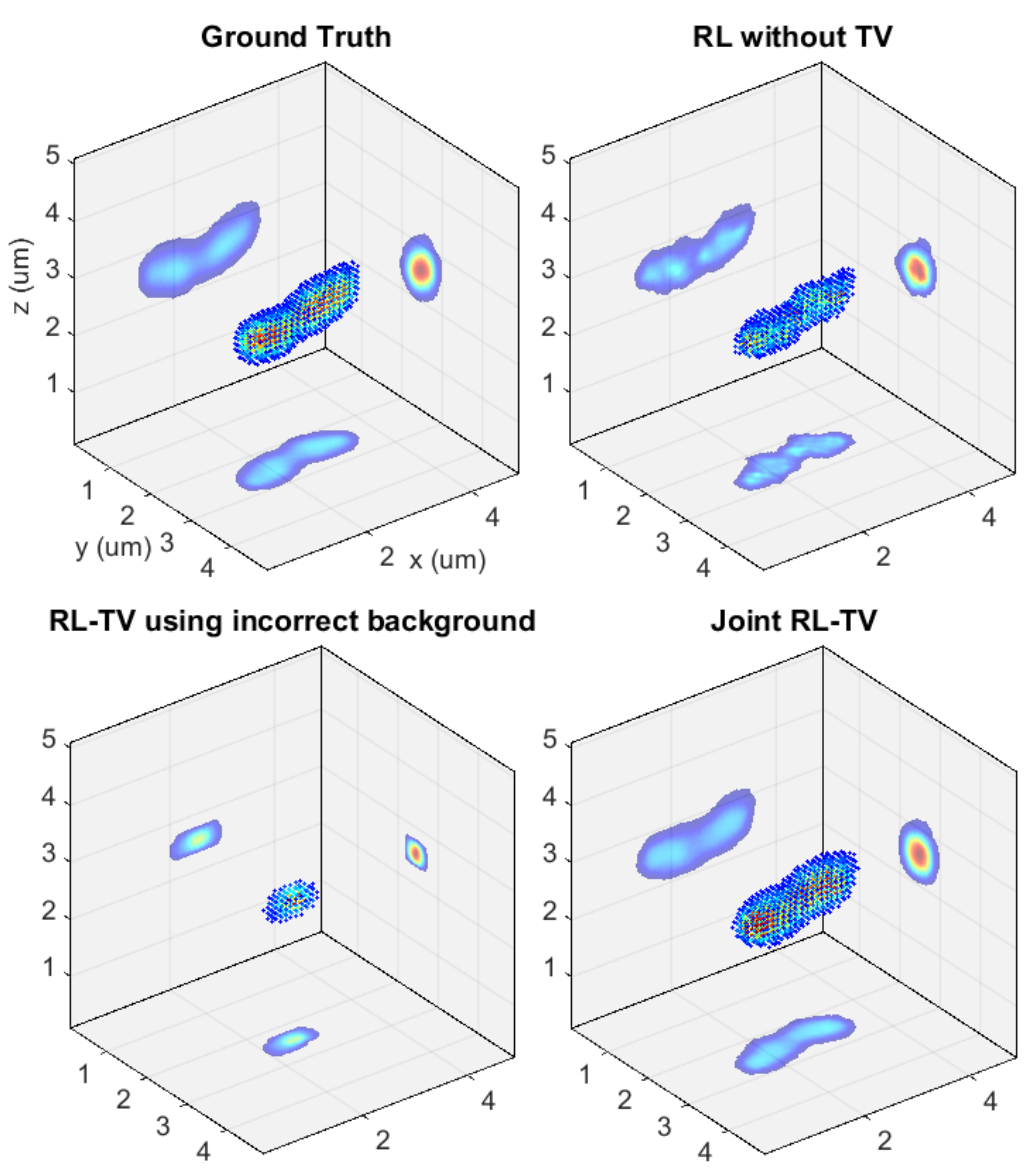}
 \caption{Simulations that demonstrate the capability of the joint RL-TV algorithm to simultaneously recover the 3D image, background noise and the optimal regularizer parameter for CMFM. Top left: a ground truth image. Top right: the standard RL deconvolution without TV regularizer ($\lambda=0$). Bottom left: RL-TV deconvolution by using an incorrect background values ($b=10$). Bottom right: joint RL-TV deconvolution can simultaneously recover a 3D image, background noise and the optimal regularizer parameter. The PSNRs for three methods are $34.2\text{dB}, 27.5\text{dB}, 40.2\text{dB}$,  and I-divergences are $204.3, 517$, and $96.7$, respectively. }
 \label{Numerical simulation}
\end{center}
\end{figure}

To verify the effectiveness of the joint regularized RL algorithm for our CMFM and also quantify the reconstruction quality, we performed a series of simulations by using an experimentally captured CMFM PSFs. 
The synthetic 3D image [Fig. \ref{Numerical simulation}(top left)] was obtained from the confocal microscope image of the 3D bacterial that was imaged under the CMFM. The CMFM measurement image was generated based on image formation model of Eq. (3)
and then degraded with background and Poisson noise. The maximum signal and background photon counts are set to be $50$ and $5$, respectively. The Poisson noise was then added to the measurement by using Matlab's Poisson random number generator.

To quantify the reconstruction quality, we computed peak signal-to-noise ratio (PSNR) and I-divergence \cite{Dey:06} between ground truth image $u$ and the reconstructed image $v$:
\begin{equation}
PSNR = 10 \log_{10}\left(\frac{\text{MAX}^2_u}{\text{MSE}}\right), \\
I_{u,v}= \sum_{i=1}^{N}\left[u_i\text{ln}\frac{u_i}{v_i} - \left(u_i - v_i \right)\right], 
\label{PSNR and I-divergence}
\end{equation}
where $\text{MAX}_u$ is the maximum pixel value of the image $u$, $\text{MSE}$ is the mean square error (MSE) between two images $u$ and $v$, and $i$ stands for the voxel index. 

\begin{figure}[htbp]
\begin{center}
 \includegraphics[width=0.6\linewidth]{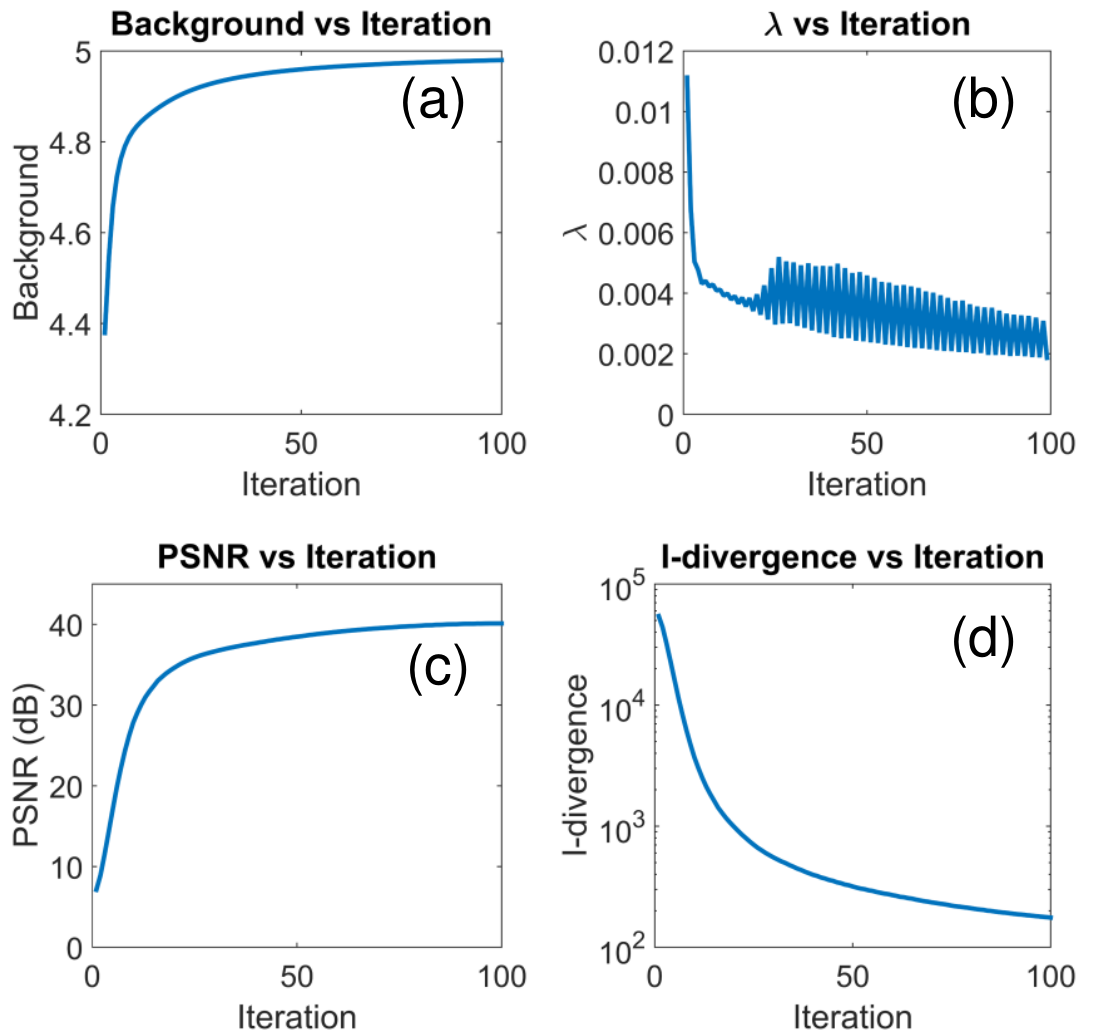}
 \caption{The convergence analysis of the joint RL-TV algorithm for the simulated CMFM. (a) The reconstructed background noise value and (b) optimal regularizer parameter during the iteration of joint RL-TV deconvolution process. (c) PSNR and (d) I-divergence between the ground truth image and reconstructed image at each iteration of the deconvolution process.}
 \label{convergence plot}
\end{center}

\section*{Appendix B: experimental results}
\label{sec:Experimental results}
We also tested the algorithm on two types of real data. The 3D reconstruction of static periplasms is shown in Fig. \ref{periplasm reconstruction}. 
The automatically recovered background noise and optimal regularizer parameter at each iteration of the joint RL-TV deconvolution process is shown in Fig. \ref{fig: background and lambda}.
\end{figure}

Figure \ref{Numerical simulation}(top right) shows the reconstruction of the standard RL deconvolution without TV constraint by setting $\lambda=0$. Because TV prior is not used, the reconstruction can not preserve the smooth edge of the original image and contains the artifacts. To investigate the effect of the background noise on the CMFM reconstruction quality, we performed RL-TV deconvolution but with an incorrect background value of $10$ (the truth value is 5). The reconstructed image is shown in Fig. \ref{Numerical simulation}(bottom left). Since the used background value is two times bigger than the real one, the recovered 3D image suffers from substantial signal loss. The reconstructed 3D image by joint RL-TV algorithm 
is shown in Fig. \ref{Numerical simulation}(bottom right). The PSNRs between the original image and reconstructed images by three methods are $34.2\text{dB}, 27.5\text{dB}, 40.2\text{dB}$,  and I-divergences are $204.3, 517$, and $96.7$, respectively. Clearly, the joint RL-TV algorithm, which simultaneously recovers 3D image, the background noise value and optimal regularizer parameter, gave the best reconstruction quality of CMFM among three methods. Figures \ref{convergence plot}(a-d) show the reconstructed background value $b$, the optimal regularizer parameter $\lambda$, PSNR and I-divergence at each iteration of the deconvolution process. Note that both the background value and regularizer parameter were initialized to be $100$, but they gradually converged to $4.99$ and $2.2\times10^{-3}$, respectively. From Fig. \ref{convergence plot}(c-d), we can also see that the image quality is dramatically improved during the iterative reconstruction process, proving the effectiveness of joint RL-TV algorithm for the 3D reconstruction of our snapshot CMFM system. 
\begin{figure}[htbp]
\centering
 \includegraphics[width=0.6\textwidth]{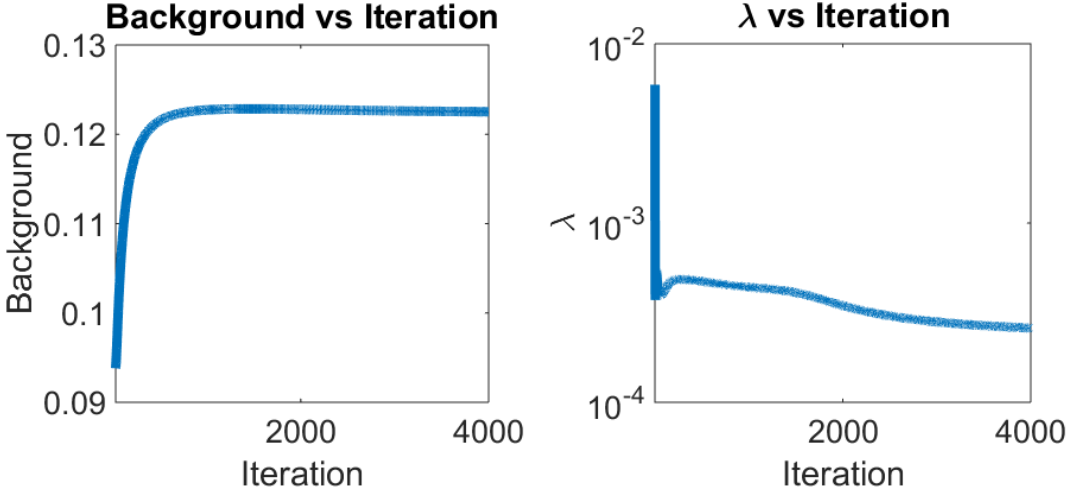}
 \caption{The recovered background values (left) and optimal regularizer parameter $\lambda$ (right) at each iteration of the joint RL-TV deconvolution process for the first experiment.}
 \label{fig: background and lambda}
\end{figure}

\begin{figure}[htbp]
\centering
 \includegraphics[width=0.6\linewidth]{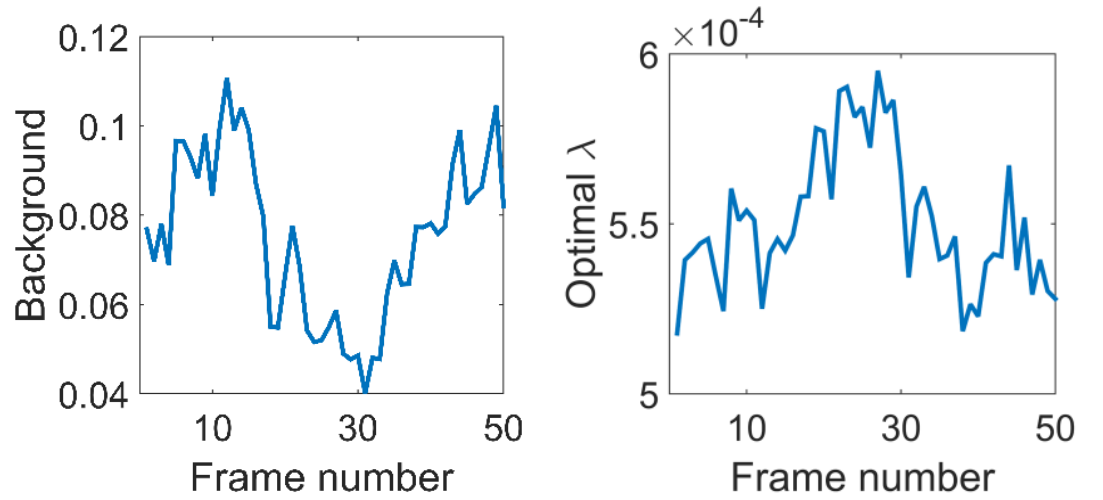}
 \caption{ The recovered background values (left) and optimal regularizer parameter $\lambda$ (right) for each video frame of a movable bacterium.}
 \label{fig: background and regularizer parameters estimation}
\end{figure}

For the second experiment, we captured an MFM video of a moving bacterium at 25 frames per second (fps) using our CMFM techniques. The complete 3D video reconstruction from the first frame to the last frame is shown in \textcolor{urlblue}{Visualization 2}. The automatically recovered background value and the optimal regularizer parameter for each video frame is shown in Fig. \ref{fig: background and regularizer parameters estimation}.

\section*{Funding}
Biological Systems Science Division, Office of Biological and Environmental Research, Office of Science, U.S. Dept. of Energy, under Contract DE-AC02-06CH11357; NSF CAREER grant IIS-1453192; ONR award N00014-15-1-2735.

\section*{Disclosures}
The authors declare that there are no conflicts of interest related to this article.



\bibliography{sample}

\begin{thebibliography}{10}
\newcommand{\enquote}[1]{``#1''}

\bibitem{Blanchard:99}
P.~M. Blanchard and A.~H. Greenaway, \enquote{Simultaneous multiplane imaging
  with a distorted diffraction grating,} {\protect\JournalTitle{Appl. Opt.}}
  \textbf{38}, 6692--6699 (1999).

\bibitem{Abrahamsson:16}
S.~Abrahamsson, R.~Ilic, J.~Wisniewski, B.~Mehl, L.~Yu, L.~Chen, M.~Davanco,
  L.~Oudjedi, J.-B. Fiche, B.~Hajj, X.~Jin, J.~Pulupa, C.~Cho, M.~Mir, M.~E.
  Beheiry, X.~Darzacq, M.~Nollmann, M.~Dahan, C.~Wu, T.~Lionnet, J.~A. Liddle,
  and C.~I. Bargmann, \enquote{Multifocus microscopy with precise color
  multi-phase diffractive optics applied in functional neuronal imaging,}
  {\protect\JournalTitle{Biomed. Opt. Express}} \textbf{7}, 855--869 (2016).

\bibitem{Oudjedi:16}
L.~Oudjedi, J.-B. Fiche, S.~Abrahamsson, L.~Mazenq, A.~Lecestre, P.-F. Calmon,
  A.~Cerf, and M.~N\"{o}llmann, \enquote{Astigmatic multifocus microscopy
  enables deep 3d super-resolved imaging,} {\protect\JournalTitle{Biomed. Opt.
  Express}} \textbf{7}, 2163--2173 (2016).

\bibitem{Abrahamsson:13}
S.~Abrahamsson, B.~H. J.~Chen, S.~Stallinga, A.~Y. Katsov, J.~Wisniewski,
  G.~Mizuguchi, P.~Soule, F.~Mueller, C.~D. Darzacq, X.~Darzacq, C.~Wu, C.~I.
  Bargmann, D.~A. Agard, M.~Dahan, , and M.~G.~L. Gustafsson, \enquote{Fast
  multicolor 3d imaging using aberration-corrected multifocus microscopy,}
  {\protect\JournalTitle{Nature methods}} \textbf{10}, 60--63 (2013).

\bibitem{Prevedel:14}
R.~Prevedel, Y.-G. Yoon, M.~Hoffmann, N.~Pak, G.~Wetzstein, S.~Kato,
  T.~Schrödel, R.~Raskar, M.~Zimmer, E.~S. Boyden, , and A.~Vaziri,
  \enquote{Simultaneous whole-animal 3d imaging of neuronal activity using
  light-field microscopy,} {\protect\JournalTitle{Nature methods}} \textbf{11},
  727--730 (2014).

\bibitem{Pegard:16}
N.~C. P\'{e}gard, H.-Y. Liu, N.~Antipa, M.~Gerlock, H.~Adesnik, and L.~Waller,
  \enquote{Compressive light-field microscopy for 3d neural activity
  recording,} {\protect\JournalTitle{Optica}} \textbf{3}, 517--524 (2016).

\bibitem{Adams:17}
J.~K. Adams, V.~Boominathan, B.~W. Avants, D.~G. Vercosa, F.~Ye, R.~G.
  Baraniuk, J.~T. Robinson, and A.~Veeraraghavan, \enquote{Single-frame 3d
  fluorescence microscopy with ultraminiature lensless flatscope,}
  {\protect\JournalTitle{Science Advances}} \textbf{3} (2017).

\bibitem{Coskun:10}
A.~F. Coskun, I.~Sencan, T.-W. Su, and A.~Ozcan, \enquote{Lensless wide-field
  fluorescent imaging on a chip using compressive decoding of sparse objects,}
  {\protect\JournalTitle{Opt. Express}} \textbf{18}, 10510--10523 (2010).

\bibitem{Antipa:18}
N.~Antipa, G.~Kuo, R.~Heckel, B.~Mildenhall, E.~Bostan, R.~Ng, and L.~Waller,
  \enquote{Diffusercam: lensless single-exposure 3d imaging,}
  {\protect\JournalTitle{Optica}} \textbf{5}, 1--9 (2018).

\bibitem{Mait:95}
J.~N. Mait, \enquote{Understanding diffractive optic design in the scalar
  domain,} {\protect\JournalTitle{J. Opt. Soc. Am. A}} \textbf{12}, 2145--2158
  (1995).

\bibitem{pedrini2001short}
G.~Pedrini and S.~Schedin, \enquote{Short coherence digital holography for 3d
  microscopy,} {\protect\JournalTitle{Optik-International Journal for Light and
  Electron Optics}} \textbf{112}, 427--432 (2001).

\bibitem{schilling1997three}
B.~W. Schilling, T.-C. Poon, G.~Indebetouw, B.~Storrie, K.~Shinoda, Y.~Suzuki,
  and M.~H. Wu, \enquote{Three-dimensional holographic fluorescence
  microscopy,} {\protect\JournalTitle{Optics Letters}} \textbf{22}, 1506--1508
  (1997).

\bibitem{wang20174d}
Z.~Wang, D.~Ryu, K.~He, O.~Cossairt, and A.~K. Katsaggelos, \enquote{4d
  tracking of biological samples using lens-free on-chip in-line holography,}
  in \emph{Digital Holography and Three-Dimensional Imaging,}  (Optical Society
  of America, 2017), pp. Tu2A--4.

\bibitem{tian20143d}
L.~Tian, J.~Wang, and L.~Waller, \enquote{3d differential phase-contrast
  microscopy with computational illumination using an led array,}
  {\protect\JournalTitle{Optics letters}} \textbf{39}, 1326--1329 (2014).

\bibitem{tian20153d}
L.~Tian and L.~Waller, \enquote{3d intensity and phase imaging from light field
  measurements in an led array microscope,} {\protect\JournalTitle{optica}}
  \textbf{2}, 104--111 (2015).

\bibitem{rosen2007digital}
J.~Rosen and G.~Brooker, \enquote{Digital spatially incoherent fresnel
  holography,} {\protect\JournalTitle{Optics letters}} \textbf{32}, 912--914
  (2007).

\bibitem{rosen2008non}
J.~Rosen and G.~Brooker, \enquote{Non-scanning motionless fluorescence
  three-dimensional holographic microscopy,} {\protect\JournalTitle{Nature
  Photonics}} \textbf{2}, 190 (2008).

\bibitem{cossairt2016compressive}
O.~Cossairt, K.~He, R.~Shang, N.~Matsuda, M.~Sharma, X.~Huang, A.~Katsaggelos,
  L.~Spinoulas, and S.~Yoo, \enquote{Compressive reconstruction for 3d
  incoherent holographic microscopy,} in \emph{Image Processing (ICIP), 2016
  IEEE International Conference on,}  (IEEE, 2016), pp. 958--962.

\bibitem{ryu2017subsampled}
D.~Ryu, Z.~Wang, K.~He, G.~Zheng, R.~Horstmeyer, and O.~Cossairt,
  \enquote{Subsampled phase retrieval for temporal resolution enhancement in
  lensless on-chip holographic video,} {\protect\JournalTitle{Biomedical optics
  express}} \textbf{8}, 1981--1995 (2017).

\bibitem{wang2017dictionary}
Z.~Wang, Q.~Dai, D.~Ryu, K.~He, R.~Horstmeyer, A.~Katsaggelos, and O.~S.
  Cossairt, \enquote{Dictionary-based phase retrieval for space-time super
  resolution using lens-free on-chip holographic video,} in \emph{Computational
  Optical Sensing and Imaging,}  (Optical Society of America, 2017), pp.
  CTu2B--3.

\bibitem{liu2017multiplexed}
H.~Y. Liu, J.~Zhong, and L.~Waller, \enquote{Multiplexed phase-space imaging
  for 3d fluorescence microscopy,} {\protect\JournalTitle{Optics Express}}
  \textbf{25}, 14986--14995 (2017).

\bibitem{wang2017compressive}
Z.~Wang, L.~Spinoulas, K.~He, L.~Tian, O.~Cossairt, A.~K. Katsaggelos, and
  H.~Chen, \enquote{Compressive holographic video,}
  {\protect\JournalTitle{Optics express}} \textbf{25}, 250--262 (2017).

\bibitem{Broxton:13}
M.~Broxton, L.~Grosenick, S.~Yang, N.~Cohen, A.~Andalman, K.~Deisseroth, and
  M.~Levoy, \enquote{Wave optics theory and 3-d deconvolution for the light
  field microscope,} {\protect\JournalTitle{Opt. Express}} \textbf{21},
  25418--25439 (2013).

\bibitem{GREEN:90}
P.~J. GREEN, \enquote{On the use of em algorithm for penalized likelihood
  estimation,} {\protect\JournalTitle{J. Royal Statist. Soc. B}} \textbf{52},
  443--452 (1990).

\bibitem{Dey:06}
N.~Dey, L.~Blanc-Feraud, C.~Zimmer, P.~Roux, Z.~Kam, J.~C. Olivo-Marin, and
  J.~Zerubia, \enquote{Richardson-lucy algorithm with total variation
  regularization for 3d confocal microscope deconvolution,}
  {\protect\JournalTitle{Microsc. Res. Tech.}} \textbf{4}, 260--266 (2006).

\bibitem{Laasmaa:11}
M.~Laasmaa, M.~Vendelin, and P.~Peterson, \enquote{Application of regularized
  richardson-lucy algorithm for deconvolution of confocal microscopy images,}
  {\protect\JournalTitle{J. Microsc}} \textbf{2}, 124--140 (2011).

\bibitem{Kuan:18}
K.~He, X.~Huang, X.~Wang, S.~Yoo, P.~Ruiz, I.~Gdor, N.~J. Ferrier, N.~Scherer,
  M.~Hereld, A.~K. Katsaggelos, and O.~Cossairt, \enquote{Design and simulation
  of a snapshot multi-focal interferometric microscope,}
  {\protect\JournalTitle{Opt. Express}} \textbf{26} (2018).

\bibitem{hugelier2016sparse}
S.~Hugelier, J.~J. De~Rooi, R.~Bernex, S.~Duw{\'e}, O.~Devos, M.~Sliwa,
  P.~Dedecker, P.~H. Eilers, and C.~Ruckebusch, \enquote{Sparse deconvolution
  of high-density super-resolution images,} {\protect\JournalTitle{Scientific
  reports}} \textbf{6}, 21413 (2016).

\bibitem{hugelier2017improved}
S.~Hugelier, P.~Eilers, O.~Devos, and C.~Ruckebusch, \enquote{Improved
  superresolution microscopy imaging by sparse deconvolution with an interframe
  penalty,} {\protect\JournalTitle{Journal of Chemometrics}} \textbf{31}, e2847
  (2017).

\bibitem{szameit2012sparsity}
A.~Szameit, Y.~Shechtman, E.~Osherovich, E.~Bullkich, P.~Sidorenko, H.~Dana,
  S.~Steiner, E.~B. Kley, S.~Gazit, T.~Cohen-Hyams \emph{et~al.},
  \enquote{Sparsity-based single-shot subwavelength coherent diffractive
  imaging,} {\protect\JournalTitle{Nature materials}} \textbf{11}, 455 (2012).

\bibitem{Solomon:18}
O.~Solomon, M.~Mutzafi, M.~Segev, and Y.~C. Eldar, \enquote{Sparsity-based
  super-resolution microscopy from correlation information,}
  {\protect\JournalTitle{Opt. Express}} \textbf{26}, 18238--18269 (2018).

\end{thebibliography}

\end{document}